\documentclass[twocolumn]{aastex631}

\usepackage{amssymb}
\usepackage{amsmath}
\usepackage[hang,flushmargin]{footmisc}
\usepackage{multirow}

\newcommand{\gv}[1]{\ensuremath{\mbox{\boldmath$ #1 $}}}

\newcommand{\dsct}[1]{$\delta$~Sct}

\shorttitle{Mode coupling in Delta Scutis}
\shortauthors{}

\begin{document}

\shortauthors{Mourabit \& Weinberg}
\shorttitle{Mode Coupling in \dsct{} stars}

\title{\bf \large Resonant Mode Coupling in $\delta$ Scuti Stars}

\correspondingauthor{Mohammed Mourabit}
\email{mohammed.mourabit@uta.edu}

\author{Mohammed Mourabit}
\affiliation{Department of Physics, University of Texas at Arlington, Arlington, TX 76019, USA}

\author[0000-0001-9194-2084]{Nevin N. Weinberg}
\affiliation{Department of Physics, University of Texas at Arlington, Arlington, TX 76019, USA}

\begin{abstract}
Delta Scuti (\dsct{}) variables are intermediate mass stars that lie at the intersection of  the main sequence and the instability strip on the Hertzsprung-Russel diagram. Various lines of evidence indicate that nonlinear mode interactions shape their oscillation spectra, including the  
 particularly compelling detection of  resonantly interacting mode triplets in the \dsct{} star KIC 8054146. Motivated by these observations, we use the theory of three-mode coupling to study the strength and prevalence of nonlinear mode interactions in fourteen \dsct{} models that span the instability strip. For each model, we calculate the frequency detunings and nonlinear coupling strengths of $\sim 10^4$ unique combinations of mode triplets.  We find that all the models contain at least $\sim 100$ well-coupled triplets whose detunings and coupling strengths are consistent with the triplets identified in  KIC 8054146. Our results suggest that resonant mode interactions can be significant in \dsct{} stars and may explain why many exhibit rapid changes in amplitude and oscillation period.
\end{abstract}

\section{\bf I\lowercase{ntroduction}}

The \dsct{} stars are pulsating variables that are on, or slightly past, the main sequence and have masses between $1.5$ and $2.5 M_\sun$, effective temperatures between 6400 and 8600 K, and stellar types between A and F  (for reviews, see \citealt{Breger:00, Goupil:05, Catelan:15, Guzik:21, Aerts:21}). Their oscillations are driven by the opacity ($\kappa$-) mechanism in the second partial ionization zone of helium and consist of low order $p$- and $g$-modes with oscillation periods ranging from $15\textrm{ min}$ to $8\textrm{ hour}$ (frequencies from a few to 100 cycles per day). The photometric amplitudes are often a few milli-magnitudes but can exceed a tenth of a magnitude in high amplitude \dsct{} stars (HADS).

The oscillation spectra from the four-year light curves of the \emph{Kepler} space mission provided unprecedented frequency resolution for over a thousand \dsct{} stars (see, e.g., \citealt{Balona:11, Uytterhoeven:11, Bowman:18}). By surveying the entire sky, the \emph{TESS} space mission has delivered month-long light curves for over ten thousand \dsct{} stars (see, e.g., \citealt{Antoci:19, BarceloForteza:20}).  Together they present an extensive asteroseismic data set that tests theoretical models of intermediate mass stars and their mode excitation. 

These space-based observations help underscore a confounding problem that was already apparent in earlier ground-based observations: \dsct{} stars with very similar global parameters often have very different oscillation frequencies and amplitudes \citep{Balona:11, Balona:18, Balona:21}. Most recently, the analysis of the oscillation spectra of 9597 \emph{TESS} \dsct{} stars by \citet{Balona:21}  finds little, if any, correlation between the frequencies and locations of the stars in the Hertzsprung-Russel diagram. He notes that otherwise similar stars can have frequency rich spectra or be dominated by just one frequency peak.  By contrast, non-adiabatic pulsation models predict that stars with similar parameters will exhibit similar oscillation spectra.  \citet{Balona:21} concludes that  an unknown mode selection process must be active and suggests that the problem may  be related to the driving of modes to nonlinear amplitudes.

Another unexplained feature of some \dsct{} stars is that their observed mode periods change much faster  than evolutionary models predict \citep{Rodriguez:95, Breger:98, Rodriguez:01, Bowman:16, Bowman:21}. For example, \citet{Breger:98} find linear period changes  $\dot{P}/P\simeq 10^{-7} \textrm{ yr}^{-1}$ in Population I (metal rich) radial pulsators, an order of magnitude faster than expected based on stellar evolution.  Moreover, they find an equal distribution of period increases and decreases whereas evolutionary models predict that they should mostly increase.  They also find that Population II (metal poor) \dsct{} stars can exhibit very sudden period changes of order $\Delta P/P \simeq 10^{-6}$.  Similarly, \citet{Bowman:21} find that the fundamental and first overtone radial modes of the HADS star KIC 5950759 exhibits a linear period change $\dot{P}/P\simeq 10^{-6} \textrm{ yr}^{-1}$ over a timescale of several years, at least two orders of magnitude larger than predicted by evolutionary models. Both \citet{Breger:98} and \citet{Bowman:21} suggest (see also \citealt{Blake:03}) that the rapid period changes could be the result of nonlinear mode interactions.  Such interactions can enable a fast transfer of energy among the modes and induce rapid, amplitude-dependent variations in their oscillation periods.

\citet{Bowman:16} find further evidence of nonlinear mode interactions in their ensemble study of \emph{Kepler} \dsct{} stars.  Of the 983 stars they analyze, 603 exhibit at least one pulsation mode that varies significantly in amplitude over four years.  While some of these amplitude variations are due to the star being in a binary, they conclude that some must be due to processes intrinsic to the star and that nonlinear mode interactions is a likely culprit. In a detailed analysis of KIC 5892969, \citet{Barcelo:15} likewise report finding amplitude variations that they argue can be attributed to nonlinear mode interactions.   Analogous to the \citet{Balona:21} study of  \emph{TESS} \dsct{} stars cited above, \citet{Bowman:16}  find  no obvious correlation between the appearance of amplitude variations and stellar parameters such as the surface gravity $\log g$ or $T_{\rm eff}$. 

Somewhat separately, \citet{Bowman:16}  note that in many \dsct{} stars the visible pulsation mode energy is not conserved over the four year Kepler observations (see also \citealt{Bowman:14}) and that low-frequency peaks can sometimes be associated with combination frequencies of high-frequency peaks.  They propose that this may be due to a transfer of energy via nonlinear mode interactions from visible modes into non-visible modes (angular degree $l\gtrsim 3$) or into the low frequency modes.

Arguably the most definitive observational evidence of nonlinear mode interactions comes from the analysis of the Kepler \dsct{} star KIC 8054146 by  \citet{Breger:14}. From the oscillation spectrum, they identify several mode triplets with properties consistent with the theory of nonlinear three-mode coupling in which two parent modes driven by the $\kappa$-mechanism nonlinearly excite a daughter mode.  In particular, for each triplet they find that: (i) the frequencies of the modes  combine such that the magnitude of their detuning $\Delta_{abc}\equiv \omega_a\pm \omega_b\pm\omega_c$ is very small ($\lesssim 10^{-5}$ times the individual frequencies), (ii) the $\textrm{O}-\textrm{C}$ phase shift of the daughter mode $c$ varies in time as $\phi_c(t)=\phi_a(t) \pm \phi_b$(t), and (iii)  the daughter's amplitude varies in proportion to the product of the parents' amplitudes, i.e.,  $A_c(t) =\mu A_a(t) A_b(t)$, where the constant $\mu$ is a measure of the nonlinear coupling strength.

There are a number of theoretical studies that consider nonlinear mode interactions in pulsating main sequence stars (e.g., \citealt{Moskalik:85, Dziembowski:85, Dziembowski:88, Moskalik:90, Buchler:97, Nowakowski:05}).  However, a comprehensive analysis investigating their impact on the oscillation modes in \dsct{} stars has not been carried out.  In this paper, we take some initial steps towards achieving such a goal.   In Section~\ref{sec:NLMC}, we present the formalism of three-mode coupling and derive the relation between the coupling strength $\mu$ and other stellar and mode parameters. In  Section~\ref{sec:calculational_methods}, we describe our calculational methods including how we search for strongly coupled modes in our \dsct{} stellar models.  In Section~\ref{sec:results}, we present the results of our analysis and compare them with the measurements of mode coupling in KIC 8054146. We summarize in Section~\ref{sec:conclusions} and discuss how future theoretical work can help further quantify the impact of nonlinear mode coupling  in \dsct{} stars.

\section{\bf R\lowercase{esonant} T\lowercase{hree}-M\lowercase{ode} C\lowercase{oupling}}
\label{sec:NLMC}

The position $\gv{x}$ of a fluid element in an unperturbed star is related to its perturbed position $\gv{x}'$ at time $t$ via $\gv{x}' = \gv{x} + \gv{\xi}({\gv{x}, t})$, where  $\gv{\xi}(\gv{x}, t)$  is  the Lagrangian displacement vector.  Since we are interested in computing weakly nonlinear mode interactions, we account for the equation of motion for $\gv{\xi}(\gv{x}, t)$ to second order in perturbation theory
\begin{align}
    \rho \ddot{\gv{\xi}} = \gv{f}_1 [\gv{\xi}] + \gv{f_2} [\gv{\xi}, \gv{\xi}],
    \label{eq:eom}
\end{align}
where $\gv{f}_1$ and $\gv{f}_2$ are the linear and second order forces, respectively (see \citealt{Schenk:02, Weinberg:12}). We solve for $\gv{\xi}(\gv{x}, t)$ by expanding the phase space vector of the displacement  in terms of its linear eigenmodes 
\begin{align}
    \begin{bmatrix}
    \gv{\xi}(\gv{x}, t)\\
    \gv{\dot{\xi}}(\gv{x}, t)
    \end{bmatrix}= \sum_a q_a(t)\begin{bmatrix}
    \gv{\xi}_a(\gv{x})\\
    -i\omega_a\gv{\xi}_a(\gv{x})
    \end{bmatrix},
    \label{eq:xi_expansion}
\end{align}
which allows us to recast the equation of motion as a set of nonlinearly coupled equations for the dimensionless amplitudes $q_a(t)$.  Here $\xi_a(\gv{x})$ and $\omega_a$ are the eigenfunctions and eigenfrequencies of a linear eigenmode and the sums are over all quantum numbers $a$ (angular degree $l_a$, azimuthal number $m_a$, and radial order $n_a$) and both frequency signs ($\pm \omega_a$).  We neglect stellar rotation in our analysis\footnote{Although some \dsct{} stars rotate rapidly (including KIC 8054146), their spin frequencies are typically much less than the frequency of the modes, with the exception, perhaps, of the low frequency $g$-modes. For example,  KIC 8054146 has a measured rotational velocity $v \sin i =300\pm 20\textrm{ km s}^{-1}$ and \citet{Breger:12} estimate that its spin frequency is $\simeq 3 \textrm{ cycles day}^{-1}$ (corresponding to $70\%$ of its Keplerian breakup frequency).  Since this is at least ten times smaller than its $p$-mode frequencies, we  do not expect rotation to drastically modify the eigenfunctions.}  and therefore the eigenfunctions are degenerate in $m_a$.  We normalize the eigenfunctions such that 
\begin{align}
    2 \omega_a^2 \int \mathrm{d}^3x \, \rho |\gv{\xi_a}|^2 = E_\star,
\end{align}
where $\rho$ is the density and $E_\star = GM^2/R$ is the characteristic energy of a star of mass $M$ and radius $R$.  The energy of a mode is then $E_a = |q_a|^2 E_\star$.

Consider a system of three coupled modes, which we label by subscripts $a$, $b$, and $c$.   If we plug the expansion given by Equation (\ref{eq:xi_expansion})  into Equation (\ref{eq:eom}), use the orthogonality of the eigenfunctions, and add a linear damping term,  we obtain an  amplitude equation for mode $a$
\begin{align}
    \dot{q}_{a} + \left(i \omega_{a}  + \gamma_{a} \right) q_{a} = i \omega_{a} \kappa_{abc} q_{b}^\ast q_{c}^\ast,
\end{align}
and similarly for modes $b$ and $c$.  Here asterisks denote complex conjugation,  $\gamma_a$ is the mode's linear damping rate (if $\gamma_a >0$) or growth rate ($\gamma_a<0$; in general $\gamma_a>0$ unless otherwise specified), and $\kappa_{abc}$ is the three-mode coupling coefficient.  The latter is dimensionless and symmetric in the three indices and is found by computing  
\begin{align}
\kappa_{abc}= \frac{1}{E_\star} \int \mathrm{d}^3x\,  \gv{\xi}_a \cdot \gv{f}_2\left[\gv{\xi}_b,\gv{\xi}_c\right].
\end{align}
We describe how we calculate $\gamma_a$ and $\kappa_{abc}$ in Section~\ref{sec:calculational_methods}. 

We are interested in a three mode system in which two of the modes are directly excited by the $\kappa$-mechanism and the third mode is excited on account of its nonlinear coupling to the other two modes (we assume it is stable to the $\kappa$-mechanism).  We refer to the former two modes  as the parents and  label them with subscripts $a$ and $b$ and we refer to the latter mode as the daughter and label it with subscript $c$. When the magnitude of the triplet's  detuning 
\begin{align}
 \Delta_{abc}=\omega_a\pm\omega_b\pm\omega_c
    \label{eq:detuning}
\end{align} 
is small, the parents can potentially drive the daughter to large amplitude. This is a type of nonlinear inhomogeneous driving that \citet{Dziembowski:82} calls direct resonance.  Another type of three-mode coupling often considered in the literature is the parametric instability, which involves one parent resonantly exciting two daughters. Although the parametric instability might also impact the observed power spectra of \dsct{} stars, in this paper we will focus on direct resonance.  This is because \citet{Breger:14} find strong evidence for direct resonance in their observations of KIC 8054146 and \citet{Bowman:16} argue it can account for the amplitude modulations of many \dsct{} stars observed by \emph{Kepler}. 

We now calculate the amplitude of a daughter mode $q_c(t)$ excited by direct resonance with two parents.  For simplicity, we assume the parent amplitudes are set entirely by the $\kappa$-mechanism and we ignore the feedback of the daughter on their amplitudes.  This should be a good approximation as long as the parent energies $E_a$ and $E_b$ are larger than the daughter's.  Since the parent energies will be nearly constant on the timescale of the oscillations\footnote{The observed amplitude modulations are on timescales of $\gtrsim 100 \textrm{ days}$ \citep{Breger:14,Bowman:16} and thus we can treat the parent amplitudes as nearly constant on the timescale of the oscillation periods ($\lesssim \textrm{ hours}$).\label{footnote:parent_amp_variations}}, we can write  $q_a(t)=A_a e^{-i\left(\omega_a t+\delta_a\right)}$, and similarly for mode $b$, where the amplitude $A_a =\sqrt{E_a/E_0}$ and the phase lag $\delta_a$ are real constants.  The daughter's amplitude equation is then just that of a driven harmonic oscillator with forcing frequency $\omega_a+\omega_b$ and driving force $i\omega_c \kappa_{abc} A_a A_b e^{i\left(\delta_a+\delta_b\right)}$.  Writing $q_c=A_c e^{i\left(\left[\omega_a + \omega_b\right]
t + \delta_c\right)}$, the daughter's amplitude equation then gives
\begin{align}
    \left(i\Delta_{abc} + \gamma_c\right)A_c e^{i\delta_c} = i \omega_c \kappa_{abc} A_a A_b e^{i\left(\delta_a + \delta_b\right)}.
    \label{eq:amp_eqn_step}
\end{align}
If we multiply each side of this equation by its complex conjugate,  we get
\begin{align}
    A_c = \mu A_a A_b,
    \label{eq:A_c}
\end{align}
where the nonlinear coupling strength
\begin{align}
      \mu = \dfrac{|\omega_c\kappa_{abc}|}{\sqrt{\Delta_{abc}^2 + \gamma_{c}^2}}.
    \label{eq:mu}
\end{align}
If instead we separate Equation~(\ref{eq:amp_eqn_step}) into its real and imaginary parts and take their ratio, we get
\begin{align}
    \tan\left(\delta_a+\delta_b\right) = \frac{\Delta_{abc} \sin \delta_c - \gamma_c \cos\delta_c}{\Delta_{abc}\cos\delta_c + \gamma_c \sin\delta_c}. 
    \label{eq:phase_lag}
\end{align}
We thus see that if  $|\Delta_{abc}| \gg \gamma_c$ then  $\delta_c \simeq \delta_a + \delta_b$, whereas if  $\gamma_c \gg |\Delta_{abc}|$ then $\delta_c\simeq \delta_a + \delta_b - \pi/2$.  This is the standard result that for a driven damped harmonic oscillator, the driving and response are in phase if the detuning is large (compared to the damping) and $\pi/2$ out of phase if the detuning is small.

\citet{Breger:14} present a similar relation for $\mu$ (their equation 5), although it looks slightly different.  This is partly because they adopt a different eigenfunction normalization.  More importantly, their expression neglects detuning and effectively assumes $|\Delta_{abc}| \ll \gamma_c$.  However, we will see that this is not generally true: for the triplets with the largest $\mu$, it is often $\Delta_{abc}$ rather than $\gamma_c$ that limits the magnitude of $\mu$.

\section{\bf C\lowercase{alculational} M\lowercase{ethods}}
\label{sec:calculational_methods}

In Section~\ref{sec:models} we describe our \dsct{} models and how we solve for their linear eigenmodes and in Section~\ref{sec:LinDam} we describe how we calculate their linear damping rates.  In Section~\ref{sec:kappa}, we explain our method for calculating the coupling coefficient $\kappa_{abc}$. In Section~\ref{sec:search_method}, we describe our procedure for finding the triplets with the largest coupling strength $\mu$ and present a combinatorics argument to understand what sets the minimum detuning. 

\subsection{Stellar Models and Linear Eigenmodes}
\label{sec:models}

\begin{figure}[t!]
\centering
\includegraphics[width=3.2in]{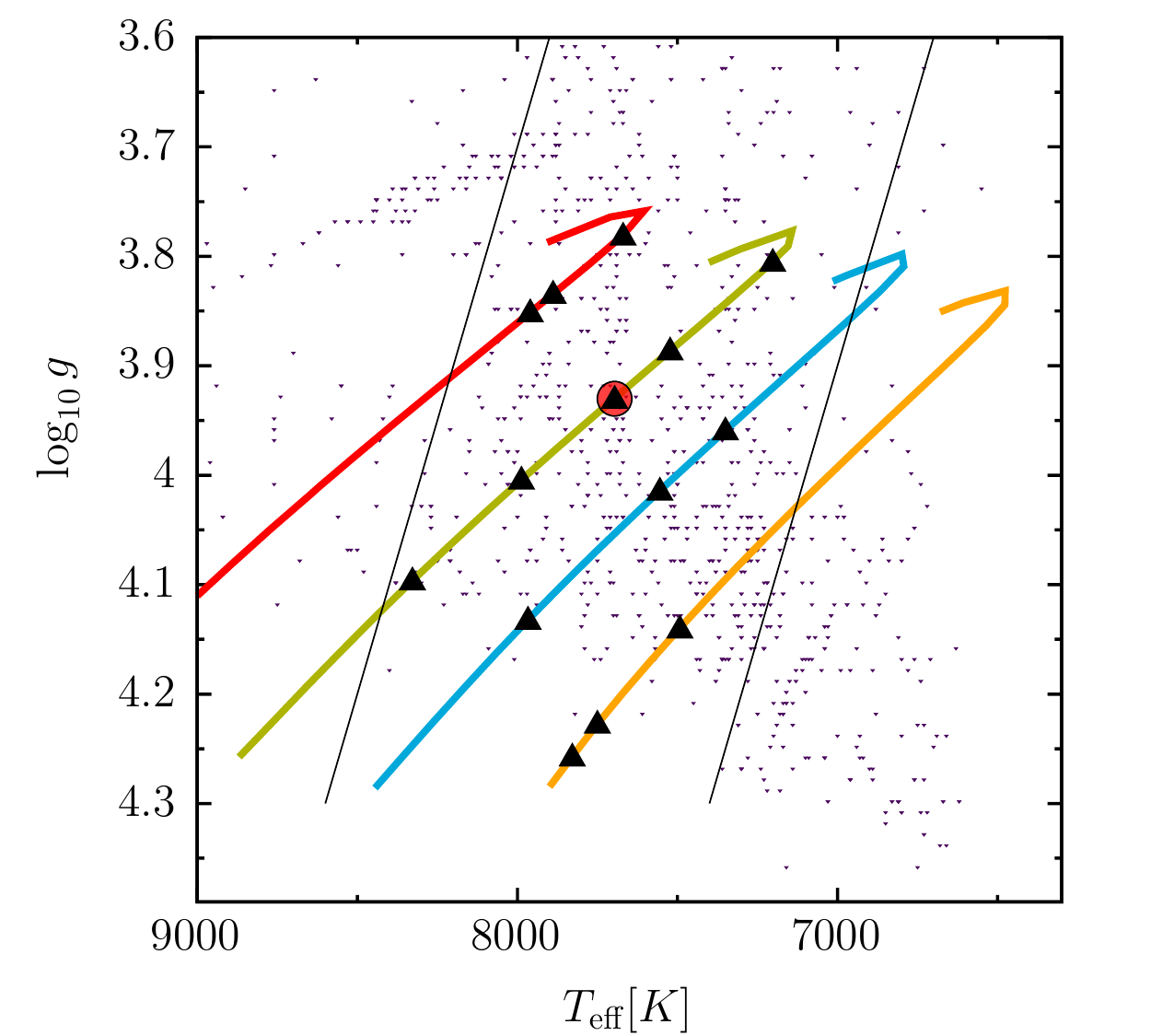}
\caption{Location of the fourteen \dsct{} stellar models (black triangles) on the $\log g$-$T_{\rm eff}$ plane, with the representative model (Section~\ref{sec:RM}) circled in red. The models have mass $M=1.7$, $1.85$, $2.0$, and $2.2 M_\sun$ and their evolution are shown by the orange, blue, green, and red curves, respectively.    The purple triangles in the background are the \emph{Kepler} \dsct{} stars from \citet{Bowman:16} and the diagonal lines are the observational blue and red edges of the classical instability strip.
\label{fig:IS}} 
\end{figure}

We use the \texttt{MESA} stellar evolution code \citep{Paxton:19} to construct fourteen \dsct{} models with masses $M=1.7$, $1.85$, $2.0$, and $2.2 M_\sun$ and effective temperatures in the range $7200 \lesssim T_{\rm eff} \lesssim 8300 \textrm{ K}$. In Figure~\ref{fig:IS}, we show the location of our models in the $\log g$--$T_{\rm eff}$ plane.    The diagonal lines show the locations of the observational blue and red edges of the instability strip from \citet{Rodriguez:01}.  We constructed our \dsct{} models so that they roughly span its range.  Figure~\ref{fig:IS} also shows the location of the \dsct{} stars observed by \emph{Kepler} \citep{Bowman:16}.

The observed power spectra of \dsct{} stars contain $g$- and $p$-modes with frequencies that range from a $\textrm{few cycles day}^{-1}$ ($g$-modes) to $\approx 100 \textrm{ cycles day}^{-1}$ ($p$-modes) and angular degree $0\leqslant l\lesssim 3$ (modes with $l\gtrsim 3$ cannot typically be resolved by \emph{Kepler}).  We use the the stellar oscillation code \texttt{GYRE} \citep{Townsend:13, Townsend:18} to find modes in this range for each of our \dsct{} models. Specifically, for each model we find the 160 modes with radial order $-20 \leqslant n \leqslant 20$  and angular degree $0 \leqslant l \leqslant 3$. For our representative model, the results of which we discuss in Section~\ref{sec:RM}, our mode frequencies range from about one to $85\textrm{ cycles day}^{-1}$. 

\begin{figure}[t!]
\centering
\includegraphics[width=2.8in]{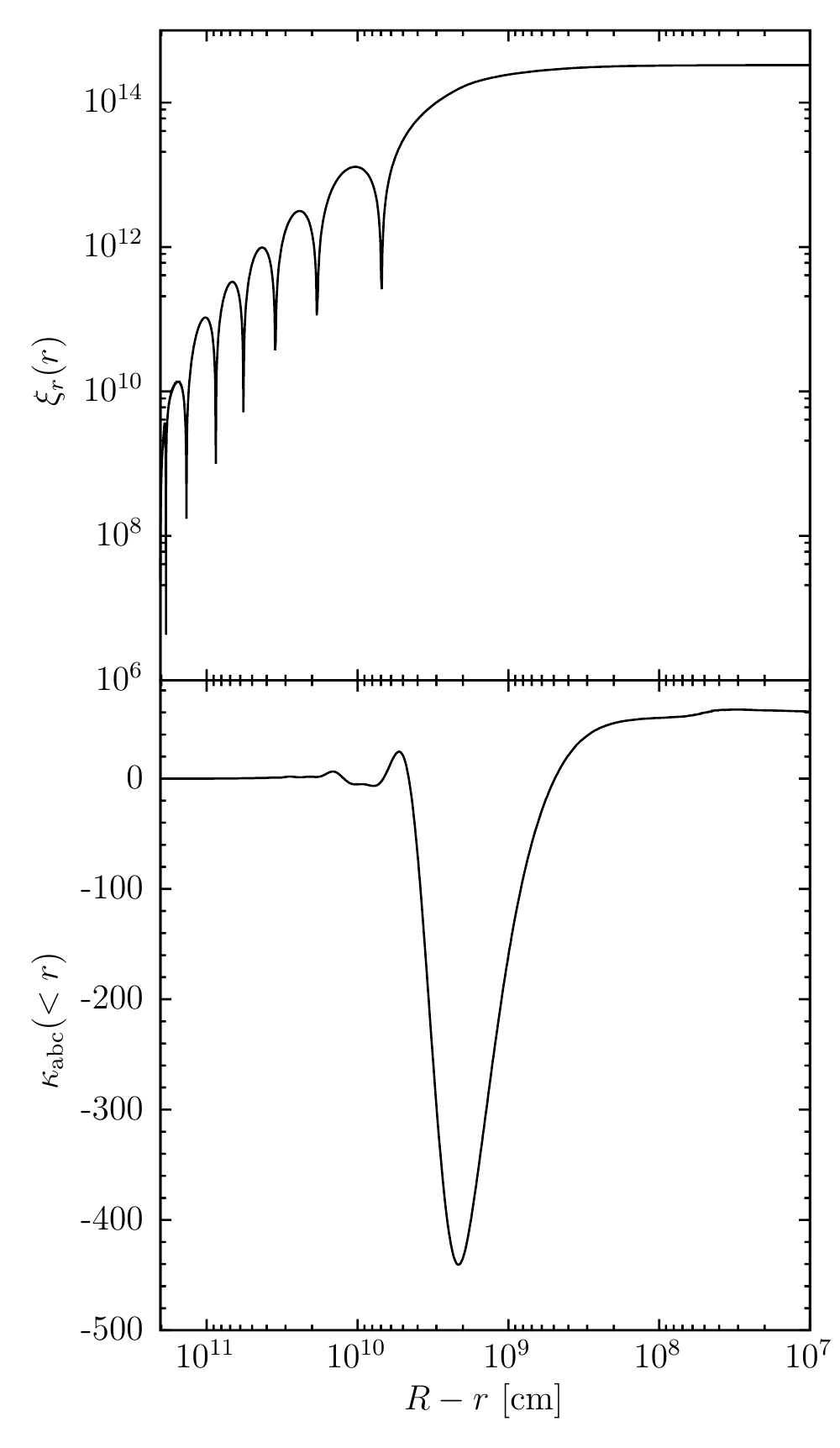}
\caption{Top panel: Radial displacement $\xi_r$ as a function of depth  $R-r$ for an $l_a=2$, $n_a=5$ $p$-mode.  Bottom panel: Cumulative integral of the coupling coefficient  $\kappa_{abc}(<r)$ for a large $\mu$ triplet (see Table~\ref{tbl:largest_per_models} in the appendix) that includes the $p$-mode shown in the top panel.  The modes of the triplet have $(l_a,l_b,l_c)=(2,2,2)$  and $(n_a, n_b, n_c) = (5, 15, 6)$ and are from the $M=2.2M_\odot$, $T_{\rm eff}=7960\textrm{ K}$ model.
\label{fig:pmode}
}
\end{figure}

\begin{figure}[t!]
\centering
\includegraphics[width=2.8in]{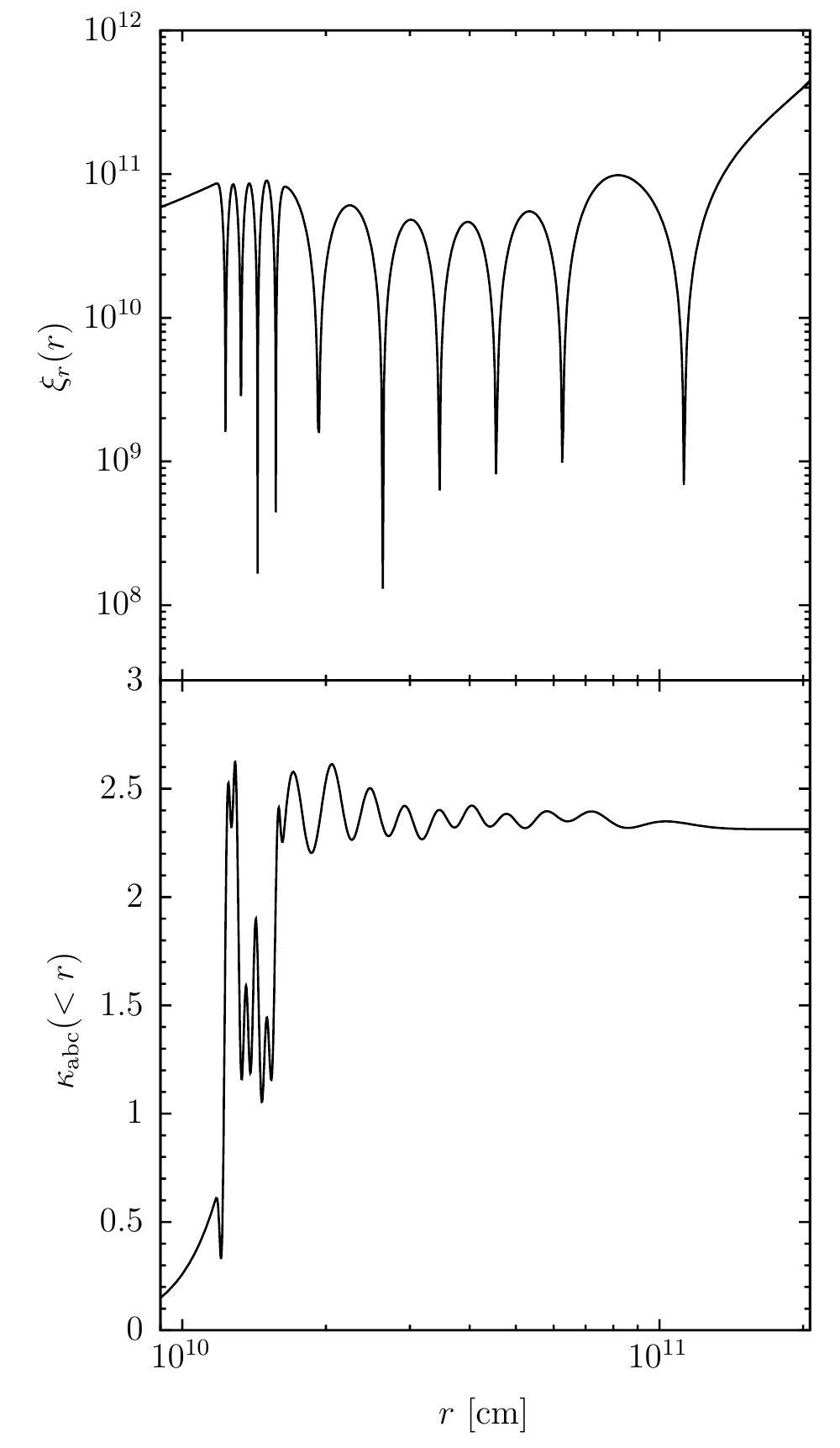}
\caption{
Top panel: Radial displacement $\xi_r$ as a function of radius for an $l_c=2$, $n_c=-10$ $g$-mode.  Bottom panel: Cumulative integral of the coupling coefficient $\kappa_{abc}(<r)$ for a large $\mu$ triplet (see  Table~\ref{tbl:largest_per_models} in the appendix) that includes the $g$-mode shown in the top panel.  The modes of the triplet have $(l_a,l_b,l_c)=(1,1,2)$ and $(n_a, n_b, n_c) = (-13, -11, -10)$ and are from the $M=2.2M_\odot$, $T_{\rm eff}=7888\textrm{ K}$ model.
\label{fig:gmode}
}
\end{figure}

In the top panels of Figures~\ref{fig:pmode} and \ref{fig:gmode} we show the radial displacement profiles $\xi_r(r)$ for a representative $p$-mode and $g$-mode, respectively. As expected, the $p$-mode amplitude peaks near the surface of the star and is orders of magnitude smaller in the  deep stellar interior (note that  the abscissa in Figure~\ref{fig:pmode} is $R-r$).  The $g$-mode amplitude, by contrast, is nearly uniform throughout the interior.  Note too that the $g$-mode's wavelength is shortest where the Brunt–Väisälä frequency peaks just outside the convective core at $r\simeq 1.0\times 10^{10}\textrm{ cm}$ and it is evanescent within it. These differences affect the spatial location of the nonlinear mode coupling within the star, as we discuss in Section~\ref{sec:kappa}.

\subsection{Linear Mode Damping}
\label{sec:LinDam}

The two principle sources of linear dissipation acting on the oscillation modes are radiative and turbulent damping.  The former is due to dissipation of the mode-induced temperature fluctuations by radiative diffusion.  The latter is due to dissipation of the mode-induced fluid displacements by turbulent eddies within the convective core. 

We calculate the radiative damping rate of a mode $\gamma_a^{(\mathrm{rad})}$ from \texttt{GYRE}'s solution of the nonadiabatic oscillation equations. For all other parts of our calculations (eigenfrequencies, displacements, etc.) we use the solution of the adiabatic oscillation equations.\footnote{This is primarily because the expressions we use to compute $\kappa_{abc}$ assume adiabatic eigenmodes.  Since the damping rates are all much smaller than the mode frequencies we consider, the modes are adiabatic to a good approximation.}  In order to assign $\gamma_a^{(\mathrm{rad})}$ to an adiabatic eigenmode, we take its values as a function of the nonadiabatic eigenfrequencies and  use interpolation to assign it to the adiabatic eigenfrequencies (the two frequencies differ only slightly).

We find the turbulent damping rate by computing
\begin{align}
    \gamma_a^{(\mathrm{turb})}=\frac{\omega_a^2}{E_\star}\int \mathrm{d}r \, \rho r^2 \nu_{\rm turb} F(r), 
\end{align}
where $F(r)$ depends on the eigenfunction displacement and is given by the expression found in \citeauthor{Higgins:68} (1968; see also \citealt{Lai:94}). The  turbulent effective viscosity $\nu_{\rm turb}$ depends on the ratio of the convective turnover frequency (provided by \texttt{MESA}) to the mode frequency and is reduced when this ratio is small.  To calculate $\nu_{\rm turb}$, we use the  power-law expression given in \citet{Duguid:20} from a fit to their numerical simulations.

\begin{figure}[t!]
\centering
\includegraphics[width=3.0in]{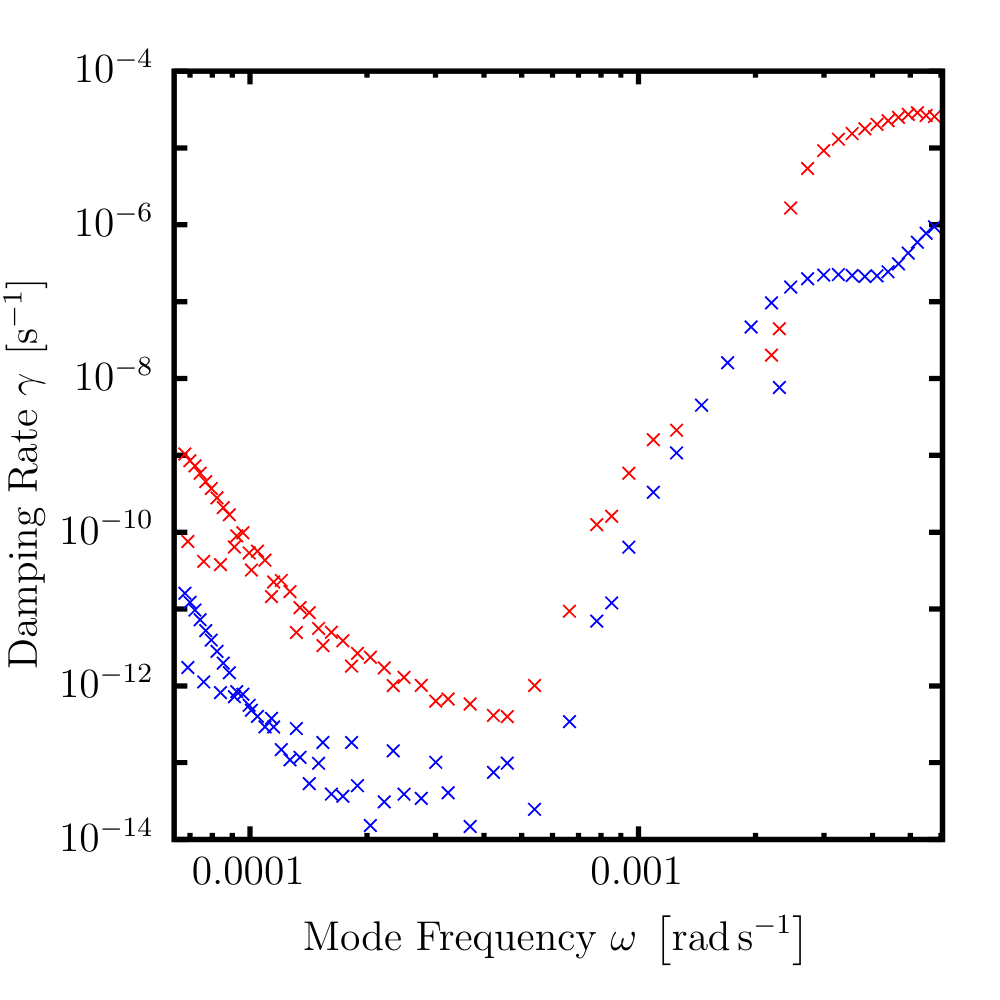}
\caption{Linear mode damping rate due to radiative (red points) and turbulent (blue points) dissipation as a function of mode frequency for the $l=2$ modes of our representative \dsct{} model ($M = 2.0 M_{\sun}$, $T_{\rm eff} = 7696 \textrm{ K}$, $\log g = 3.9$).
\label{fig:Dampings}}
\end{figure}

In Figure~\ref{fig:Dampings} we show $\gamma_a^{(\mathrm{rad})}$ and $\gamma_a^{(\mathrm{turb})}$ as a function of mode frequency for all the $l=2$ modes of the $2.0M_\sun,7696\textrm{ K}$ model (the other models yield similar results). We see that the radiative damping dominates at essentially all frequencies, often by a factor of $\sim 100$, and therefore the total damping rate 
\begin{align}
\gamma_a^{(\rm tot)} = \gamma_a^{(\mathrm{rad})} + \gamma_a^{(\mathrm{turb})} \simeq \gamma_a^{(\mathrm{rad})}.    
\end{align}
The $p$-mode damping rates are several orders of magnitude larger than the $g$-mode damping rates because the $p$-modes have much smaller mode inertias (see \citealt{Aerts:10}).  Moreover, since the efficiency of radiative dissipation increases with decreasing wavelength, we see that $\gamma_a^{(\mathrm{rad})}$ increases with increasing  $p$-mode frequency and decreasing $g$-mode frequency.

\subsection{Nonlinear coupling coefficient $\kappa_{abc}$ }
\label{sec:kappa}

We calculate $\kappa_{\mathrm{abc}}$ using Equations A55 through A62 in \citet{Weinberg:12}.
The modes couple only if they satisfy the angular selection rules $|\ell_b-\ell_c|\le \ell_a \le \ell_b+\ell_c$ with $\ell_a+\ell_b+\ell_c$ even and $m_a+m_b+m_c=0$.   The angular selection rules help restrict the number of   triplets to consider in our search for large $\mu$ values. In Section~\ref{sec:results} we show that the triplets with the largest $\mu$ have $\kappa_{abc} \sim 1 - 100$.  

As representative examples of the $\kappa_{abc}$ calculation, in Figure~\ref{fig:pmode} we show the cumulative integral $\kappa_{abc}(<r)=\int_0^r  (\mathrm{d}\kappa_{\mathrm{abc}}/\mathrm{d}r) \, \mathrm{d}r$ for a triplet of three $p$-modes and in Figure~\ref{fig:gmode} for a triplet of three $g$-modes.  For the triplet shown in Figure~\ref{fig:pmode}, $\kappa_{abc}\simeq 60$ and most of the contribution to $\kappa_{abc}$ (i.e., most of the nonlinear coupling) occurs in the outer layers of the star.  This triplet has 
fractional detuning $\Delta_{abc}/\omega_c\simeq 1.3\times10^{-4}$, $\gamma_c= 9.9 \times 10^{-8} \textrm{ s}^{-1}$, and $\mu= 3.7 \times 10^5$. For the triplet shown in Figure~\ref{fig:gmode}, $\kappa_{abc}\simeq 2.3$ and most of the contribution occurs in the deep interior just outside the convective-radiative interface.  This is because that is where the mode amplitudes of the respective triplets peak. This triplet has
$\Delta_{abc}/\omega_c\simeq 3.0 \times 10^{-7}$, $\gamma_c= 1.0 \times 10^{-12}\textrm{ s}^{-1}$, and $\mu= 7.6\times10^6$.

\begin{figure*}
\centering
\includegraphics[width=7.0in]{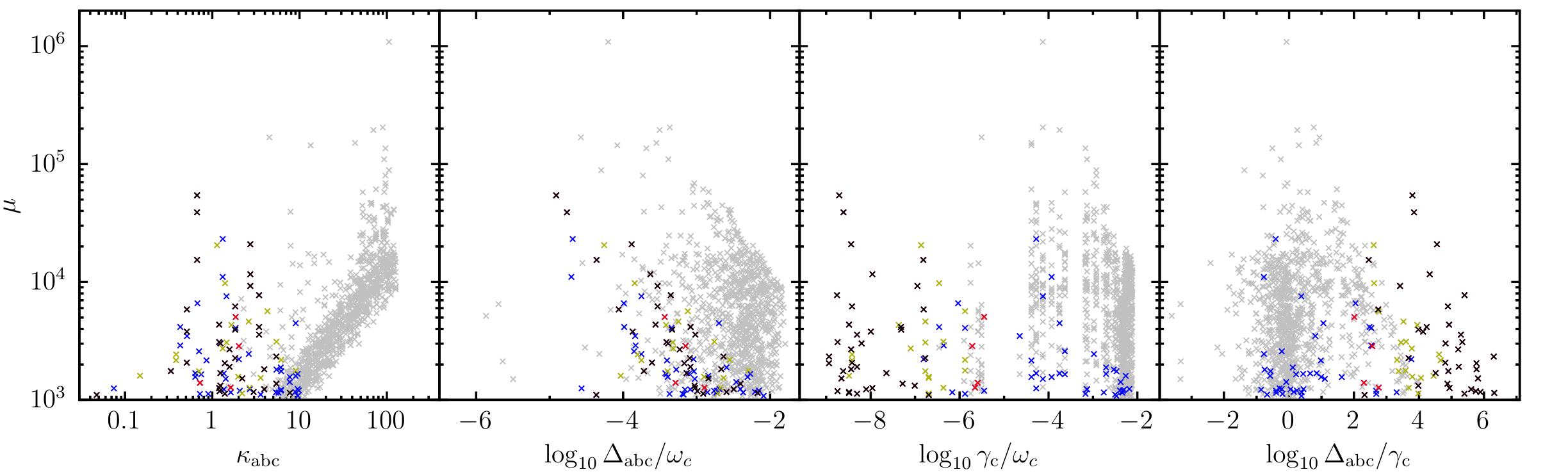}
\caption{Triplets with nonlinear coupling strength $\mu > 10^3$ from the $M= 2.0 M_\sun$, $T_{\rm eff} = 7696 \textrm{ K}$, model.  The panels show $\mu$ as a function of the coupling coefficient $\kappa_{abc}$, the fractional detuning $\Delta_{abc}/\omega_c$, the daughter damping rate $\gamma_c$ (in units of $\omega_c$), and the ratio $\Delta_{abc}/\gamma_c$.  The different colors correspond to different triplet types: three $p$-modes (grey points), three $g$-modes (black points), two $p$-mode parents and a $g$-mode daughter (green points), two $g$-mode parents and a $p$-mode daughter (red points), one parent a $p$-mode and the other a $g$-mode (blue points; the daughter can be either a $p$- or $g$-mode).
\label{fig:fiduciary}}
\end{figure*}

\subsection{Search for triplets with large  coupling strength $\mu$}
\label{sec:search_method}

We are interested in finding the triplets with the largest values of the nonlinear coupling strength $\mu$.  From Equation (\ref{eq:mu}), we see this requires finding triplets with large $\kappa_{abc}$ and small $\Delta_{abc}/\omega_c$ and $\gamma_c/\omega_c$. For each of our \dsct{} models, we search for the triplets with the largest $\mu$ by scanning over all combinations of mode triplets among our sample of 160 modes. Since computing $\kappa_{abc}$ is the most expensive part of the $\mu$ calculation, we restrict our search to triplets with detuning $|\Delta_{abc}| < 0.15 \sqrt{GM/R^3}$. In practice, this does not affect our results since the largest $\mu$ all have much smaller detunings than this.  As we show in Section~\ref{sec:min_detuning},  the magnitude of the smallest detunings can be understood as resulting from the repeated drawing of three random numbers from a uniform distribution.  Note that when calculating $\Delta_{abc}$, we must account for all combinations of mode frequency signs (see Equation~\ref{eq:detuning}) since the phase space mode expansion includes both positive and negative frequencies for each $\omega_a$.  This means that the daughters are not necessarily the highest frequency mode of the triplet; indeed, for many of our largest $\mu$ they are the intermediate frequency mode (\citealt{Breger:12} find this as well among the triplets they identify in KIC 8054146).

\subsubsection{Minimum detuning $\Delta_{abc}$}
\label{sec:min_detuning}

In Section~\ref{sec:results} we show that all of our \dsct{} models have minimum fractional detunings  $\Delta_{abc}/\omega_c \approx 10^{-5}$.  We can roughly understand what sets this minimum as follows.  For each model, the mode frequencies lie between some minimum and maximum, $\omega_{\rm min} \le \omega_a \le \omega_{\rm max}$, corresponding to the highest order $g$- and $p$-modes. To a reasonable approximation, we can treat the 160 calculated modes from each model as though their frequencies are uniformly distributed between these two values.\footnote{Since the modes are relatively low-order, they do not satisfy the asymptotic relations appropriate for high-order modes, in which the period ($g$-modes) or frequency ($p$-modes) spacings are nearly constant for fixed $l$.}
Assume, therefore, that $a\equiv \left(\omega_a-\omega_{\rm min}\right)/\left(\omega_{\rm max}-\omega_{\rm min}\right)$ is uniformly distributed  between $0$ and $1$, and write $\Delta_{abc}=\omega_a+\omega_b-\omega_c$ as 
$\epsilon = a+b-c$, where $\epsilon\equiv  \left(\Delta_{abc}-\omega_{\rm min}\right)/\left(\omega_{\rm max}-\omega_{\rm min}\right)$ is similar to the fractional detuning $\Delta_{abc}/\omega_c$. For a random draw of $\{a,b,c\}$, the probability  $P(|\epsilon| < \epsilon_0) = \epsilon_0 - \epsilon_0^2/4\simeq \epsilon_0$ for $0<\epsilon_0\ll 1$.\footnote{An approximate way to understand this is that there is a $50\%$ chance $a+b<1$.  If it is, there is a $2 \epsilon_0$ chance that $c$ will be within $\epsilon_0$ of $a+b$ and therefore $P(|\epsilon| < \epsilon_0) \simeq \epsilon_0$.}  
On average, we therefore need $\simeq 1/\epsilon_0$ draws of $\{a,b,c\}$ before we get an $|\epsilon| < \epsilon_0$.   For $0\le l \le 3$, there are eight combinations of $\{l_a,l_b,l_c\}$ that satisfy the selection rules (see Section~\ref{sec:kappa}) and since $-20\le n \le 20$, there are about ${40 \choose 3} \simeq 10^4$ combinations of $\{n_a,n_b,n_c\}$. Our search therefore has $\approx 10^5$ unique combinations of $\{a,b,c\}$ and we can  expect a minimum $|\epsilon| \approx 10^{-5}$, consistent with our calculated minimum $\Delta_{abc}/\omega_c$.

\citet{Breger:14} find that the triplets in KIC 8054146 have fractional detunings $\lesssim 10^{-6}$, somewhat smaller than our calculated minimum.  This could be because we do not account for rotation, which lifts the degeneracy in $m_a$.  It therefore increases the total number of unique triplet combinations and further reduces the minimum  $\Delta_{abc}/\omega_c$.

\section{\bf R\lowercase{esults}}
\label{sec:results}

In Section~\ref{sec:RM}, we show results for a representative \dsct{} model with $M=2.0 M_\sun$, $T_{\rm eff}= 7696\textrm{ K}$, and $\log g=3.9$.  We choose this model because it lies near the middle of the instability strip (see Figure~\ref{fig:IS}) and because KIC 8054146 has similar $T_{\rm eff}$ and $\log g$ \citep{Breger:12, Breger:14}.  In Section~\ref{sec:other models}, we show  results for our other \dsct{} models, which we find are generally similar to those of our representative model. In Section~\ref{sec:KIC8054146} 
we compare our results to the triplets identified in KIC 8054146.

\subsection{Representative \dsct{} model}
\label{sec:RM}

In Figure \ref{fig:fiduciary}, we show the triplets with the largest coupling strengths $\mu$ from our representative model ($M=2.0 M_\sun$, $T_{\rm eff}= 7696\textrm{ K}$, $\log g=3.9$). We see that there are many triplets with $\mu > 10^4$ and even a few with $\mu > 10^5$. The four panels, from left to right, show how $\mu$ depends on the coupling coefficient $\kappa_{abc}$, the fractional detuning $\Delta_{abc}/\omega_c$, the daughter damping rate $\gamma_c/\omega_c$, and $\Delta_{abc}/\gamma_c$.  Although the coupling strength can be as large as $\mu \sim 10^4$  even for coupling coefficients as small as $\kappa_{abc}\approx 1$, the largest $\mu$ tend to have $\kappa_{abc} \gtrsim 10$. These are generally the triplets containing three $p$-modes (grey points). Nonetheless, other types of triplets involving different combinations of $p$- and $g$-modes (colored points) can still have $\mu \gtrsim 10^3$. The second panel shows that  the largest $\mu$ are more likely to have small detunings, with $\Delta_{abc}/\omega_c \lesssim 10^{-3} $ (in  Section~\ref{sec:min_detuning} we explain why the minimum $\Delta_{abc}/\omega_c \approx 10^{-5}$).  By contrast, the third panel shows that the largest $\mu$ are not especially sensitive to $\gamma_c$; if anything they are more likely to have larger $\gamma_c$ (those are the ones in which the daughter is a $p$-mode; see Figure~\ref{fig:Dampings}).  This is because for the majority of strongly coupled triplets $\Delta_{abc} \gtrsim \gamma_c$, as can be seen in the fourth panel. Thus, by Equation (\ref{eq:mu}),  $\mu$ tends to be limited by the magnitude of $\Delta_{abc}$ rather than $\gamma_c$.

\begin{figure}[t!]
\centering
\includegraphics[width=3.2in]{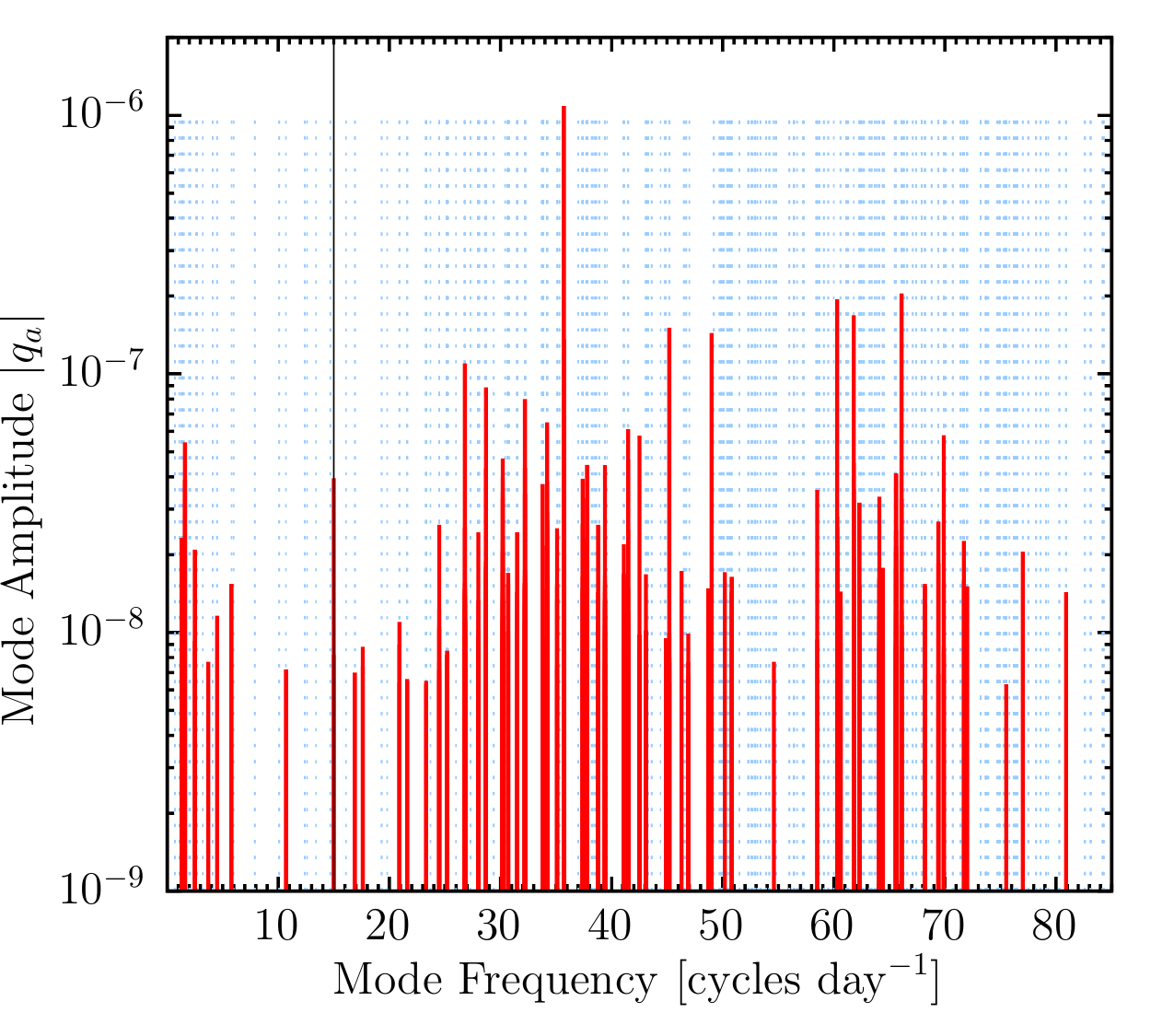}
\caption{Mode amplitude as a function of frequency for triplets of the $M = 2.0 M_{\sun}$, $T_{\rm eff} = 7696\textrm{ K}$ model. The parents (blue lines) are assumed to be at a fixed amplitude $A_a=A_b=10^{-6}$.  The daughters (red lines) are at amplitude $A_c=\mu A_a A_b$. 
 Only triplets with $\mu > 10^{3}$ are shown.  The black line indicates the dynamical frequency $\sqrt{GM/R^3}$ which approximately separates the $p$- and $g$-modes.
\label{fig:PS}}
\end{figure}

In Figure~\ref{fig:PS} we show (in red) the amplitude of the daughter modes $A_c = |q_c|$ as a function of their frequency for the representative \dsct{} model.  This figure is like a power-spectrum, albeit a fairly artificial  one.  Specifically, to calculate $A_c$ we use Equation~(\ref{eq:A_c}) assuming the parents (in blue) are all at amplitude $A_a=A_b=10^{-6}$.  While this choice of parent amplitude is essentially arbitrary (the actual amplitudes are set by the $\kappa$-mechanism driving of the parents, which we do not attempt to model), the reason we choose $A_a=A_b=10^{-6}$ is because then the best coupled triplets ($\mu \sim 10^5$) have $A_c\sim 10^{-7}$ to $10^{-6}$.  Thus, our choice ensures the daughter amplitudes are comparable to their parents' amplitudes, which is similar to the daughter modes observed in KIC 8054146 \citep{Breger:14}.  Interestingly, we see in Figure~\ref{fig:PS} that while most of the high amplitude daughters are $p$-modes (frequencies greater than about $15\textrm{ cycles day}^{-1}$; see black vertical line), the $g$-mode daughters can also have significant  amplitudes.  This too is a feature observed in KIC 8054146.

A notable difference between our artificial power spectrum (Fig.~\ref{fig:PS}) and the observed power spectrum of KIC 8054146 (see Fig.~1 in \citealt{Breger:14}) is that ours has a much higher density (per unit frequency) of parent and daughter modes with large amplitudes.   Although we do not know for certain, we suspect that this is because in our treatment we assume every $p$- and $g$-mode parent is unstable to the $\kappa$-mechanism and linearly driven to $A_a=A_b=10^{-6}$.  In reality, as the power spectrum of KIC 8054146 indicates, only a subset of these modes will be linearly unstable  and driven to sufficiently large amplitudes to be detectable\footnote{Note, however, that \citet{Breger:14} only analyze the twenty dominant modes and their harmonics. The full spectrum presented in \citet{Breger:12} contains many more modes.}  (and to likewise drive daughters to detectable amplitudes). 

In order to determine which modes are unstable to the $\kappa$-mechnaism, we could turn to  \texttt{GYRE}'s solution of the nonadiabatic oscillation equations and find which modes have $\gamma_a^{(\rm rad)} < 0$.  \citet{Goldstein:20} use this approach to find the unstable modes in models of $\beta$ Cephei stars whose oscillations are driven by the iron-bump $\kappa$-mechanism.  However, in practice we find that only low-order modes with $n\approx\textrm{few}$ have $\gamma_a^{(\rm rad)} < 0$. Moreover, when we include turbulent dissipation $\gamma_{a}^{(\rm turb)}$, which \texttt{GYRE} does not account for, many of these modes have total damping $\gamma_{a}^{(\rm tot)} > 0$.  Compared to the observed spectra of \dsct{} stars, \texttt{GYRE} seems to find too few  unstable modes and the range of frequencies is too low ($f \lesssim 40\textrm{ cycles day}^{-1}$).  The origin of this discrepancy is unclear, though it could be related to the larger problem of the unknown mode selection process noted in the introduction.

This important caveat aside, we find two triplets whose  parents are both unstable according to \texttt{GYRE} ($\gamma_{a}^{(\rm tot)} < 0$) and whose $\mu$ is  large enough ($\mu \simeq 10^4$) to drive a daughter to significant amplitude (again assuming  $A_a=A_b=10^{-6}$). These two triplets are listed in Table~\ref{tbl:largest_per_models} in the appendix (see the lines starting with asterisks among the triplets of the representative model).

\subsection{Our other \dsct{} models}
\label{sec:other models}

\begin{figure}[t!]
\centering
\includegraphics[width=3.2in]{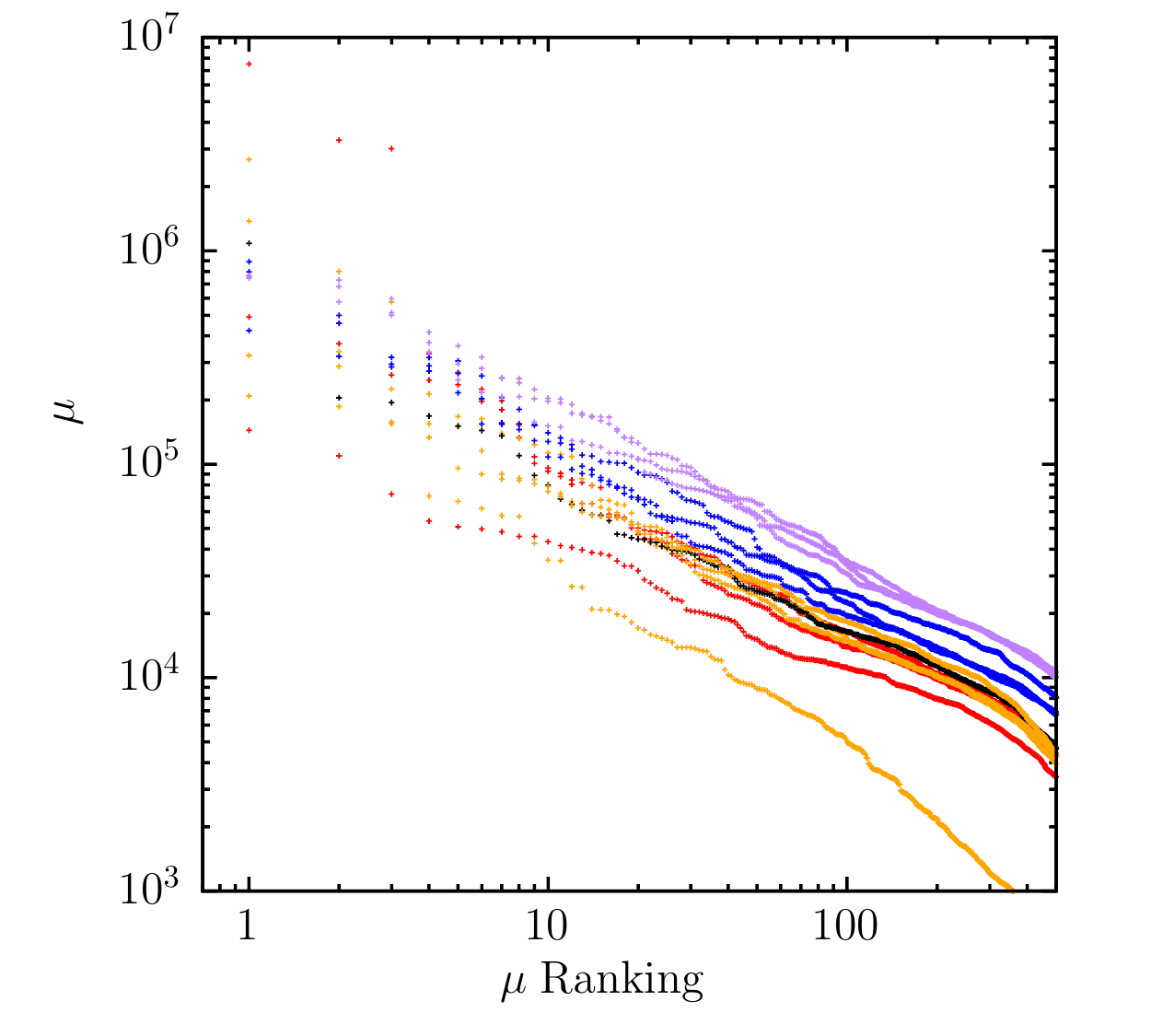}
\caption{Rank ordering of triplets according to their nonlinear coupling strength $\mu$ for the fourteen \dsct{} models.  The points are colored according to the stellar mass: $M = 2.2 M_{\sun}$ (red), $M = 2.0 M_{\sun}$ (orange  with the representative model in black), $M = 1.85 M_{\sun}$ (blue), and $M = 1.7 M_{\sun}$ (purple). 
\label{fig:largest}}
\end{figure}

In Figure~\ref{fig:largest} we show the largest $\mu$ triplets in rank order for each of our fourteen \dsct{} models.  We find that all the models, save one, have more than one hundred triplets with $\mu>10^4$ and a few triplets with $\mu > 10^5$. Triplets with strong nonlinear coupling are thus a common feature of our \dsct{} models.

\begin{figure}[t!]
\centering
\includegraphics[width=3.6in]{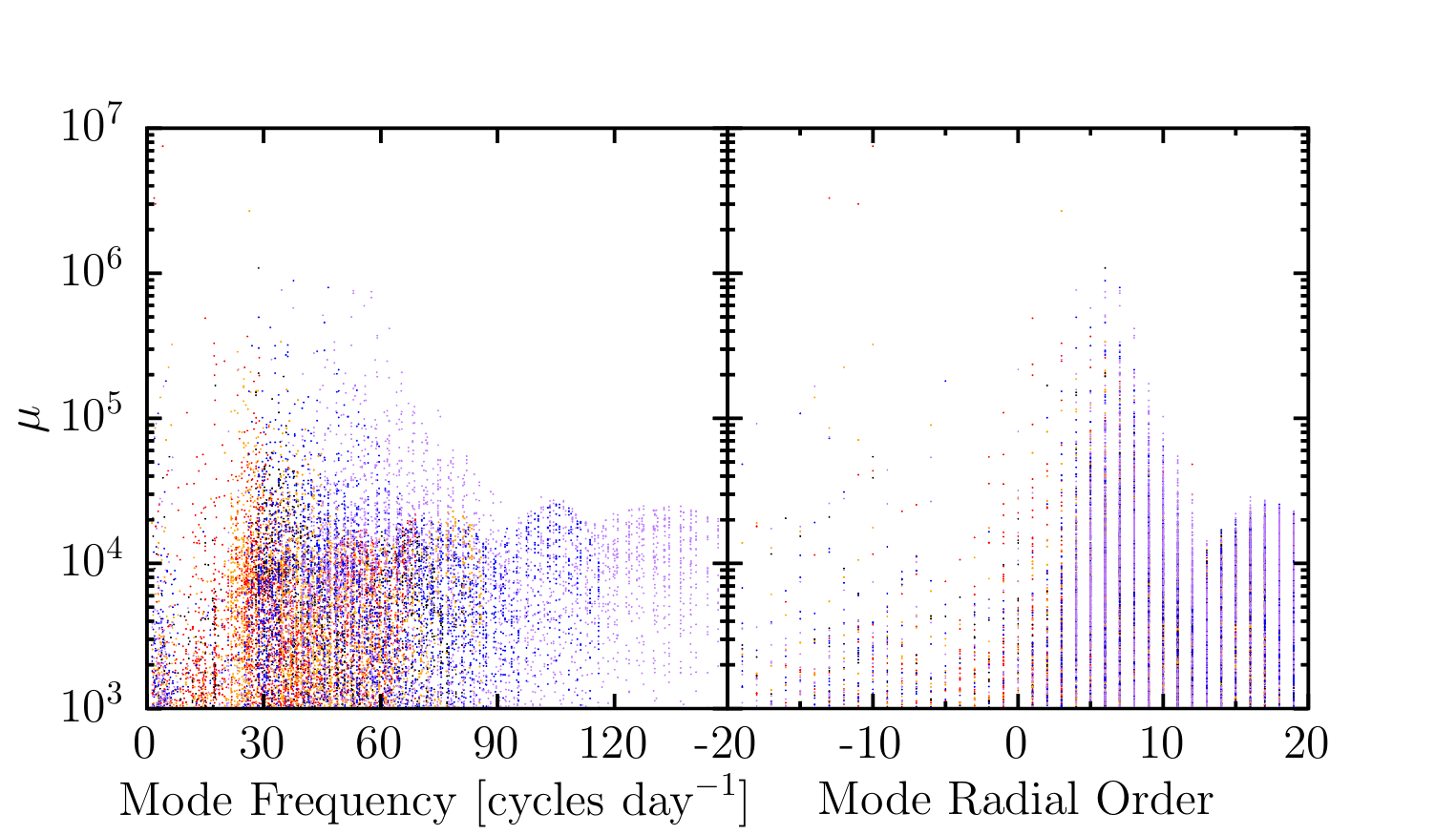}
\caption{Nonlinear coupling strength $\mu$ as a function of daughter mode frequency (left panel) and radial order (right panel) for the fourteen \dsct{} models.  The points are colored according to the stellar mass (as in Fig.~\ref{fig:largest}): $M = 2.2 M_{\sun}$ (red), $M = 2.0 M_{\sun}$ (orange  with the representative model in black), $M = 1.85 M_{\sun}$ (blue), and $M = 1.7 M_{\sun}$ (purple). 
\label{fig:mu_omega}}
\end{figure}

In Table~\ref{tbl:largest_per_models} in the appendix we provide more detailed information about the triplets with the three largest $\mu$ for each model. Most of these triplets consist of three $p$-modes, although some consist of three $g$-modes or a mix of $p$- and $g$-modes. With only a few exceptions,  the daughter is either the lowest frequency or intermediate frequency mode of the triplet. While high-frequency daughters tend not to be among the highest-$\mu$ triplets (those with $\mu \sim 10^4-10^5$), Figure~\ref{fig:mu_omega} shows that the triplets with $\mu\sim 10^3$ often do contain high-frequency,  high-order, $p$-mode daughters ($f \gtrsim 60 \textrm{ cycles day}^{-1}$, $n\gtrsim 10$).  It also shows that they sometimes contain low-frequency  $g$-mode daughters ($f \lesssim 15 \textrm{ cycles day}^{-1}$, $n < 0$) despite their somewhat smaller $\kappa_{abc}$ (see  Fig.~\ref{fig:fiduciary}).

As with the representative model, we also list in Table~\ref{tbl:largest_per_models}  the two largest $\mu$ triplets whose parents are both unstable, i.e., $\gamma_{a}^{(\rm tot)} < 0$ according to \texttt{GYRE} (not all models have such triplets).  In general, they consist of low-order modes and have $\mu \gtrsim 10^3$.

\subsection{Comparison to the \dsct{} star KIC 8054146}
\label{sec:KIC8054146} 

The stellar parameters of our representative \dsct{} model  ($M=2.0 M_\sun$, $T_{\rm eff}=7696\textrm{ K}$, and $\log g=3.9$) are similar to those of the Kepler \dsct{} star KIC 8054146 studied by \citet{Breger:14}.  They found several resonant triplets in their analysis of the star's power spectrum, and focus on three of them in particular.  Their frequencies $(f_a,f_b,f_c)$ in units of $\textrm{ cycles day}^{-1}$ are $(25.9509, 60.4346, 34.4836)$, $(25.9509, 63.3680, 37.4170)$, and $(25.9509, 66.2988, 40.3479)$.  All three share a common mode ($f_a=25.9509 \textrm{ cycles day}^{-1}$) and to the precision provided in the paper they have   detunings $|f_a-f_b+f_c| < 10^{-4} \textrm{ cycles day}^{-1}$ corresponding to fractional detunings $|\Delta_{abc}|/f_c < 3\times10^{-6}$ (given the four year data set, the frequencies cannot be measured to better than $\sim 10^{-4}\textrm{ cycles day}^{-1}$ and only an upperbound can be placed on the detunings). They find that the amplitudes  vary over the four-year Kepler observations according to Equation~(\ref{eq:A_c}). This allows them to identify the parent and daughter modes  in each triplet  since they are observed to vary in concert as $A_c(t) = \mu A_a(t) A_b(t)$.  Consistent with the mode coupling interpretation, they also find that their phase shifts vary as $\phi_c(t)=\phi_a(t)\pm \phi_b(t)$.  The daughter is the intermediate frequency mode in all three triplets (mode $f_c$ in the list above) and from its amplitude variation they measure a coupling strength in the range $7000 \lesssim \mu\lesssim 60,000$. 

This range of measured $\mu$ is consistent with the values we find in our calculations (not counting the few triplets we find with $\mu \gtrsim 10^5$). Moreover, as can be seen from Table~\ref{tbl:largest_per_models} in the the appendix, our representative \dsct{} model contains triplets whose other properties are similar to those found in KIC 8054146, especically the  triplet with the third largest $\mu$ in the list.  Specifically, their frequencies and detunings are similar and since $|\Delta_{abc}| > \gamma_c$,  we expect their phase shift, like the phase lag, to satisfy $\phi_c \simeq \phi_a\pm \phi_b$ (see Section~\ref{sec:NLMC}). Some of the triplets in  KIC 8054146 also contain and a mix of $p$- and $g$-modes, like ours. However, given that we  find hundreds of triplets with $\mu \gtrsim 10^4$, this similarity between the largest $\mu$ triplets listed in Table~\ref{tbl:largest_per_models} and those observed in KIC 8054146 could just be a coincidence.  As we discuss below, a more definitive comparison requires a time-dependent mode network calculation that also models the parent driving.

\section{\bf S\lowercase{ummary} \lowercase{and} C\lowercase{onclusions}}
\label{sec:conclusions}

Motivated by the observational evidence of nonlinear mode interactions in \dsct{} stars, especially in KIC 8054146,  we carried out a theoretical investigation of the prevalence and strength of resonant three-mode coupling in  fourteen models of \dsct{} stars that span the instability strip.  For each model, we found all the eigenmodes with angular degree $0\le l\le 3$ and radial order $-20\le n \le 20$, corresponding to $g$- and $p$-modes with frequencies in the range $1 \lesssim f \lesssim 100 \textrm{ cycles day}^{-1}$.  We computed the linear damping rate $\gamma$ of each mode due to radiative and turbulent dissipation and found that the former typically dominates.  We then  searched for all mode triplets $(a,b,c)$ that satisfy the three-mode angular selection rules and computed their detuning $\Delta_{abc}=\omega_a\pm \omega_b\pm \omega_c$ and coupling coefficient $\kappa_{abc}$. Lastly, we rank-ordered the triplets according to their nonlinear coupling strength $\mu=|\omega_c \kappa_{abc}|/\sqrt{\Delta_{abc}^2+\gamma_c^2}$.  

According to the theory of nonlinear three-mode coupling, two parent modes $a$ and $b$ driven by the $\kappa$-mechanism to amplitudes $A_a$ and $A_b$ will excite a daughter mode $c$ to an amplitude $A_c=\mu A_a A_b$. In all of our models, we found at least ten triplets with $\mu \gtrsim 10^4$, and in many models there were $> 100$ such triplets. These $\mu$ values are broadly consistent with those directly measured by \citet{Breger:14} in their analysis of KIC 8054146.  We found that the triplets with large $\mu$ consist of various combinations of $p$- and $g$-modes (e.g., three $p$-modes, three $g$-modes, two $p$-mode parents and a $g$-mode daughter), which is also true of the triplets found in KIC 8054146.

Our results suggest that resonant three-mode interactions can be significant in \dsct{} stars, and the ubiquity of large $\mu$ across all of our models may explain why amplitude variations \citep{Bowman:16} and large $\dot{P}/P$ \citep{Breger:98, Blake:03, Bowman:21} are commonly observed. However, our analysis did not model the parent driving and therefore did not solve for the overall mode amplitudes. Thus, we cannot yet say the extent to which mode coupling impacts the observed oscillation spectra. Moreover, we only focused on direct nonlinear forcing, in which two parent modes excite a daughter mode. Another form of three-mode coupling that might impact the oscillation spectra is  the parametric instability, in which a parent mode excites two daughter modes (see, e.g., \citealt{Dziembowski:82}). Unlike direct forcing, the parametric instability is only triggered if a parent is above a threshold amplitude.  \citet{Dziembowski:85} showed that it is a potentially important mechanism for limiting oscillation amplitudes and transferring energy across the modes (see also \citealt{ Dziembowski:88}). A further complication is that the daughters can themselves excite granddaughters and the granddaughters can excite great-granddaughters etc., either through direct forcing, the parametric instability, or both.

In order to make further progress on this problem, a time-dependent mode network calculation is needed.  This would entail solving a large set of  nonlinearly coupled amplitude equations that account for the parent driving, multiple generations of $p$- and $g$-modes, and both forms of three-mode coupling.  \citeauthor{Weinberg:21} (2021; see also \citealt{Weinberg:19}) carried out a similar calculation in their study of solar-like oscillations in red giants, although there the parents were driven by stochastic turbulent motions not the $\kappa$-mechanism.  They found that depending on the evolutionary state of the red giant, the secondary modes (daughters, granddaughters, etc.) could significantly suppress the parent amplitudes and efficiently transfer the parents' energy among thousands of higher-order, higher-degree modes.  The structure  of a red giant, and therefore the mode coupling,  is of course very different from that of a \dsct{} star.  Most notably, the nonlinear mode interactions in a red giant are concentrated in the stellar core and involve mixed modes. In a \dsct{} star, by contrast, we found that many of the triplets consist of three $p$-modes whose nonlinear couplings peak near the stellar surface.   Such surface interactions might impact the oscillation spectra more directly and leave a detectable, time-dependent imprint on not just the mode amplitudes, but also on their frequencies and line widths.\\

This work was supported by NASA ATP grant 80NSSC21K0493. We thank Phil Arras and Hang Yu for valuable discussions.

\software{\texttt{MESA } \citep[][\url{http://mesa.sourceforge.net}]{Paxton:19},
\texttt{GYRE } \citep[][\url{https://gyre.readthedocs.io/en/stable/}]{Townsend:13, Townsend:18}
}

\appendix

\section{Table of triplets with large nonlinear coupling strength}
\begin{center}
\begin{longtable*}{r r r r r r r r r r r r  r r r}
   \caption{Triplets with the largest values of the nonlinear coupling strength $\mu$ from each \dsct{} model.    The subscripts $a$ and $b$ label the parents and subscript $c$ labels the daughter. The columns are: angular degree $l$, radial order $n$, mode frequency $f$ (in cycles day$^{-1}$), total damping rate (in units of $\log_{10} \gamma^{(\rm tot)} / \omega_c$), frequency detuning (in units of $\log_{10} \Delta_{abc}/\omega_c$), coupling coefficient $\kappa_{abc}$, and $\log_{10} \mu$. For each model, the first three rows show the largest $\mu$ triplets.  The following two rows (indicated by asterisks) show the   largest $\mu$ triplets whose  parents both have negative damping rates (we then give $\log_{10} [-\gamma_a^{(\rm tot)}/\omega_c]$; note that not all models have such triplets).  Two of the triplets containing three $g$-modes repeat with different parent-daughter combinations, and we only list the one with the largest $\mu$.
   \label{tbl:largest_per_models}}
   \\
   \hline
      $l_a$ & $l_b$ & $l_c$ & $n_a$ & $n_b$ & $n_c$ & $f_a$ & $f_b$ & $f_c$ & $ \log \gamma_a$ & $ \log \gamma_b$  & $ \log \gamma_c$ &  $\log \Delta_{\rm abc}$ & $\kappa_{\mathrm{abc}}$ & $\log \mu$\\ 
   \hline \hline
   \multicolumn{15}{c}{ $M = 2.2  M_{\odot}$, $\hspace{0.2cm} T_{\rm eff} = 7670 \textrm{ K}$, $\hspace{0.2cm} \log g = 3.78$} \\
   \hline  
   3 & 3 & 0 & 19 & 10 & 7 & 61 & 38 & 23 & -1.9 & -2.0 & -3.5 & -4.6 & -46.3 & 5.2\\
   0 & 1 & 1 & 17 & 13 & -1 & 52 & 41 & 10 & -1.6 & -1.5 & -6.1 & -4.7 & 2.1 & 5.0\\
   0 & 3 & 3 & 8 & 15 & 5 & 26 & 50 & 24 & -2.7 & -1.9 & -3.0 & -4.5 & 77.8 & 4.9\\
   * 3 & 2 & 1 & 2 & 4 & 12 & 19 & 20 & 39 & -5.0 & -5.3 & -2.2 & -2.2 & 49.7 & 3.7\\
   * 1 & 2 & 1 & 4 & 4 & 12 & 18 & 20 & 39 & -5.2 & -5.3 & -2.2 & -3.1 & 34.5 & 3.7\\
   [0ex] \hline
   \multicolumn{15}{c}{ $M = 2.2  M_{\odot}$, $\hspace{0.2cm}  T_{\rm eff} = 7888 \textrm{ K}$, $\hspace{0.2cm} \log g = 3.83 $} \\
   \hline
   1 & 1 & 2 & -13 & -11 & -10 & 2 & 2 & 4 & -7.1 & -6.6 & -8.5 & -6.5 & 2.3 & 6.9\\
   3 & 2 & 1 & 4 & 11 & 3 & 23 & 40 & 17 & -5.0 & -1.9 & -4.8 & -4.2 & 21.5 & 5.5\\
   2 & 1 & 1 & 6 & 13 & 3 & 28 & 45 & 17 & -3.1 & -1.8 & -4.8 & -3.9 & 30.7 & 5.4\\
   * 3 & 2 & 1 & 4 & 4 & 13 & 23 & 22 & 45 & -5.4 & -5.1 & -2.2 & -2.7 & 64.5 & 4.0\\
   * 0 & 1 & 1 & 6 & 5 & -14 & 23 & 21 & 2 & -3.7 & -4.3 & -6.0 & -2.1 & 0.8 & 2.0\\
   [0ex]
   \hline
   \multicolumn{15}{c}{ $M = 2.2  M_{\odot}$, \hspace{0.2cm}  $T_{\rm eff} = 7960 \textrm{ K}$, \hspace{0.2cm} $\log g = 3.85$}\\
   \hline
   0 & 2 & 2 & 6 & 10 & 1 & 23 & 38 & 15 & -4.9 & -2.0 & -4.9 & -4.8 & 9.3 & 5.7\\
   2 & 2 & 2 & 6 & 15 & 5 & 29 & 55 & 26 & -3.7 & -2.0 & -4.3 & -3.8 & 59.8 & 5.6\\
   2 & 2 & 2 & 15 & 5 & 6 & 55 & 26 & 29 & -2.0 & -4.3 & -3.7 & -3.9 & 59.8 & 5.4\\
   * 3 & 1 & 2 & 4 & 6 & 13 & 24 & 25 & 48 & -5.5 & -6.9 & -2.2 & -2.4 & 75.2 & 4.0\\
   * 3 & 2 & 1 & 4 & 4 & 13 & 24 & 23 & 47 & -5.5 & -5.6 & -2.2 & -2.5 & 63.0 & 3.9\\
   [0ex]
    \hline
   \multicolumn{15}{c}{ $M = 2.0  M_{\odot}$, \hspace{0.2cm}  $T_{\rm eff} = 7202 \textrm{ K}$, \hspace{0.2cm} $\log g = 3.80$}\\
   \hline 
   3 & 0 & 3 & 4 & 4 & -10 & 23 & 17 & 6.5 & -4.7 & -4.6 & -8.8 & -6.5 & -0.1 & 5.5\\
   1 & 0 & 1 & 7 & 15 & 6 & 26 & 49 & 23 & -2.7 & -1.8 & -3.5 & -3.7 & 109.0 & 5.5\\
   1 & 2 & 3 & -13 & -14 & -12 & 2 & 3 & 6 & -6.4 & -8.4 & -8.6 & -5.3 & -1.1 & 5.4\\
   * 0 & 2 & 2 & 3 & -2 & -17 & 14 & 11 & 3 & -4.6 & -5.7 & -8.0 & -4.5 & 0.4 & 4.1\\
   * 3 & 1 & 2 & 0 & 3 & 8 & 16 & 17 & 33 & -5.7 & -5.3 & -2.5 & -2.7 & 6.5 & 3.3\\
   [0ex] \hline
   \multicolumn{15}{c}{ $M = 2.0  M_{\odot}$, \hspace{0.2cm}  $T_{\rm eff} = 7523 \textrm{ K}$, \hspace{0.2cm} $\log g = 3.89$}\\
   \hline 
   2 & 1 & 1 & 16 & 9 & 6 & 63 & 37 & 27 & -1.8 & -2.4 & -3.8 & -3.5 & 77.7 & 5.3\\
   2 & 2 & 2 & 8 & 15 & 4 & 35 & 60 & 25 & -2.8 & -1.7 & -4.3 & -3.5 & 57.4 & 5.3\\
   2 & 2 & 0 & 17 & 9 & 7 & 67 & 38 & 29 & -1.8 & -2.4 & -3.4 & -4.5 & 64.2 & 5.2\\
   * 2 & 1 & 1 & 2 & 4 & -15 & 19 & 20 & 2 & -4.5 & -4.2 & -7.5 & -3.5 & 0.5 & 3.2\\
   * 3 & 0 & 3 & 0 & 5 & 8 & 17 & 22 & 40 & -7.5 & -5.7 & -3.7 & -2.3 & 4.6 & 3.0\\
   [0ex] \hline
   \multicolumn{15}{c}{ $M = 2.0  M_{\odot}$, \hspace{0.2cm}  $T_{\rm eff} = 7696 \textrm{ K}$, \hspace{0.2cm} $\log g = 3.93$}\\
   \hline 
   0 & 1 & 1 & 16 & 8 & 6 & 64 & 36 & 29 & -1.8 & -2.8 & -4.1 & -4.2 & 105.7 & 6.0\\
   2 & 1 & 1 & 8 & 16 & 6 & 37 & 66 & 29 & -2.6 & -1.8 & -4.1 & -3.4 & 89.6 & 5.3\\
   1 & 2 & 3 & 5 & 14 & 6 & 25 & 60 & 35 & -5.3 & -2.0 & -3.8 & -3.5 & 70.0 & 5.3\\
   * 3 & 3 & 0 & 3 & 3 & 12 & 25 & 25 & 49 & -5.4 & -5.4 & -2.3 & -2.6 & 30.4 & 3.7\\
   * 3 & 3 & 2 & 3 & 3 & 11 & 25 & 25 & 49 & -5.4 & -5.4 & -2.4 & -2.4 & 18.7 & 3.5\\
   [0ex]
   \hline
   \multicolumn{15}{c}{ $M = 2.0  M_{\odot}$, \hspace{0.2cm}  $T_{\rm eff} = 7987 \textrm{ K}$, \hspace{0.2cm} $\log g = 4.00$}\\
   \hline
   0 & 2 & 2 & 17 & 10 & 3 & 77 & 51 & 26 & -1.8 & -2.1 & -5.2 & -4.8 & 44.6 & 6.4\\
   1 & 1 & 2 & 17 & 9 & 6 & 79 & 45 & 34 & -2.0 & -2.7 & -4.4 & -3.6 & 84.3 & 5.5\\
   2 & 3 & 3 & 7 & 15 & 6 & 38 & 74 & 36 & -2.9 & -1.9 & -3.6 & -3.4 & 72.3 & 5.2\\
   * 3 & 2 & 1 & 4 & 5 & 13 & 32 & 31 & 62 & -5.3 & -5.4 & -2.2 & -2.1 & 51.9 & 3.7\\
   * 3 & 3 & 2 & 4 & 4 & 13 & 32 & 32 & 64 & -5.3 & -5.3 & -2.2 & -2.4 & 38.9 & 3.7\\
   [0ex]\hline
   \multicolumn{15}{c}{ $M = 2.0  M_{\odot}$, \hspace{0.2cm}  $T_{\rm eff} = 8328 \textrm{ K}$, \hspace{0.2cm} $\log g = 4.10$}\\
   \hline
   1 & 1 & 2 & -13 & -18 & -13 & 2 & 1 & 3 & -8.0 & -7.0 & -7.9 & -5.8 & 2.4 & 6.1\\
   1 & 2 & 3 & 15 & 5 & 7 & 82 & 36 & 47 & -2.0 & -4.8 & -3.5 & -3.0 & 70.4 & 4.9\\
   1 & 3 & 2 & 15 & 8 & 4 & 82 & 52 & 31 & -1.8 & -3.0 & -4.6 & -3.2 & 46.7 & 4.8\\
   [0ex]
   \hline
   \multicolumn{15}{c}{ $M = 1.85  M_{\odot}$, \hspace{0.2cm}  $T_{\rm eff} = 7350 \textrm{ K}$, \hspace{0.2cm} $\log g = 3.96$}\\
   \hline
   0 & 3 & 3 & 7 & 15 & 6 & 33 & 71 & 38 & -3.4 & -1.9 & -3.9 & -4.3 & 132.7 & 5.9\\
   2 & 0 & 2 & 17 & 11 & 4 & 77 & 49 & 29 & -1.8 & -2.2 & -4.0 & -3.9 & 73.1 & 5.7\\
   3 & 3 & 0 & 6 & 15 & 7 & 38 & 71 & 33 & -3.8 & -1.8 & -3.4 & -4.2 & 132.7 & 5.5\\
   * 0 & 2 & 2 & 2 & -1 & 4 & 15 & 14 & 29 & -6.6 & -7.1 & -4.0 & -3.4 & 1.3 & 3.5\\
   * 1 & 2 & 3 & 1 & 1 & 5 & 15 & 19 & 34 & -6.5 & -6.2 & -3.8 & -3.5 & 1.1 & 3.5\\
   [0ex]
   \hline
   \multicolumn{15}{c}{ $M = 1.85  M_{\odot}$, \hspace{0.2cm}  $T_{\rm eff} = 7555 \textrm{ K}$, \hspace{0.2cm} $\log g = 4.01$}\\
   \hline
   1 & 1 & 2 & 19 & 12 & 5 & 92 & 60 & 32 & -1.7 & -2.0 & -4.3 & -4.3 & 31.2 & 5.6\\
   0 & 0 & 0 & 18 & 10 & 7 & 85 & 49 & 36 & -1.8 & -2.4 & -3.5 & -4.3 & 102.7 & 5.5\\
   3 & 3 & 0 & 16 & 8 & 7 & 82 & 46 & 36 & -1.8 & -2.5 & -3.5 & -3.6 & 121.2 & 5.5\\
   * 3 & 3 & 0 & 2 & 2 & 10 & 25 & 25 & 49 & -6.7 & -6.7 & -2.5 & -2.1 & 5.3 & 2.8\\
   * 0 & 0 & 0 & 3 & 4 & 9 & 20 & 24 & 45 & -7.1 & -6.4 & -2.7 & -4.0 & 1.0 & 2.7\\
   [2ex]
   \hline
   \multicolumn{15}{c}{ $M = 1.85  M_{\odot}$, \hspace{0.2cm}  $T_{\rm eff} = 7967 \textrm{ K}$, \hspace{0.2cm} $\log g = 4.13$}\\
   \hline
   2 & 1 & 1 & 8 & 17 & 7 & 54 & 100 & 47 & -3.0 & -1.9 & -3.8 & -4.3 & 123.4 & 5.9\\
   3 & 0 & 3 & 18 & 11 & 6 & 111 & 65 & 46 & -2.0 & -2.4 & -4.0 & -3.9 & 80.6 & 5.7\\
   3 & 0 & 3 & 19 & 12 & 6 & 116 & 70 & 46 & -2.0 & -2.2 & -4.0 & -4.3 & 35.7 & 5.5\\
   [0ex]
   \hline
   \multicolumn{15}{c}{ $M = 1.7  M_{\odot}$, \hspace{0.2cm}  $T_{\rm eff} = 7492 \textrm{ K}$, \hspace{0.2cm} $\log g = 4.14$}\\
   \hline
   3 & 3 & 2 & 17 & 11 & 4 & 109 & 75 & 35 & -1.7 & -2.1 & -5.0 & -4.4 & 30.3 & 5.9\\
   1 & 2 & 1 & 17 & 10 & 5 & 104 & 67 & 38 & -1.8 & -2.3 & -4.5 & -3.9 & 70.0 & 5.8\\
   1 & 1 & 2 & 16 & 8 & 6 & 99 & 53 & 45 & -1.9 & -3.0 & -3.6 & -4.1 & 122.3 & 5.7\\
   * 2 & 1 & 1 & 2 & -1 & -2 & 27 & 16 & 11 & -7.2 & -7.5 & -9.0 & -4.1 & -0.4 & 3.7\\
   * 2 & 2 & 2 & -2 & 1 & 5 & 15 & 25 & 40 & -8.9 & -6.8 & -4.2 & -4.1 & 0.4 & 3.6\\
   [0ex]
   \hline
   \multicolumn{15}{c}{ $M = 1.7  M_{\odot}$, \hspace{0.2cm}  $T_{\rm eff} = 7750 \textrm{ K}$, \hspace{0.2cm} $\log g = 4.23$}\\
   \hline
   0 & 0 & 0 & 19 & 11 & 7 & 131 & 78 & 53 & -1.8 & -2.4 & -3.8 & -4.2 & 120.1 & 5.9\\
   3 & 3 & 0 & 18 & 10 & 7 & 133 & 80 & 53 & -1.8 & -2.4 & -3.8 & -4.3 & 113.4 & 5.9\\
   2 & 0 & 2 & 7 & 16 & 6 & 59 & 111 & 53 & -3.3 & -1.9 & -3.9 & -3.7 & 111.4 & 5.7\\
   \hline 
   \multicolumn{15}{c}{ $M = 1.7  M_{\odot}$, \hspace{0.2cm}  $T_{\rm eff} = 7829 \textrm{ K}$, \hspace{0.2cm} $\log g = 4.26$}\\
   [0ex]
    \hline
   3 & 0 & 3 & 18 & 11 & 6 & 140 & 82 & 58 & -1.8 & -2.5 & -3.8 & -4.3 & 129.5 & 5.9\\
   2 & 1 & 3 & 16 & 8 & 6 & 123 & 65 & 58 & -1.9 & -3.2 & -3.8 & -4.3 & 118.1 & 5.8\\
   3 & 3 & 0 & 19 & 11 & 7 & 147 & 91 & 56 & -1.8 & -2.3 & -3.9 & -4.0 & 87.8 & 5.8\\
   [0ex]
   \hline
   \end{longtable*}
\end{center}

\bibliography{refs}

\end{document}